\documentstyle[11pt,newpasp,twoside,epsf]{article}
\def\gapprox{{_>\atop{^\sim}}}
\def\lapprox{{_<\atop{^\sim}}}

\def\cmmd{\rm {cm^{-3}}}

\def\s-1{\rm {s^{-1}}}
\def\twco{$^{12}$CO}
\def\thco{$^{13}$CO}

\def\etal {et al.}
\def\kms {\hbox{${\rm km\,s}^{-1}$}}
\def\Msun{\ifmmode\hbox{M}\sun\else M$\sun$\fi}
\def\Mgs{\ifmmode M_{\rm gas}\else $M_{\rm gas}^{*}$\fi}
\def \r{${\cal R}_{1-0}$}
\def\sun{\ifmmode_{\mathord\odot}\else$_{\mathord\odot}$\fi}
\def\Lsun{\ifmmode\hbox{L}\sun\else L$\sun$\fi}

\def\edcomment#1{\iffalse\marginpar{\raggedright\sl#1\/}\else\relax\fi}
\reversemarginpar

\begin{document}

\title{Gas Properties in the Medusa Minor Merger --- Comparing with ULIRGs}
\author{S. Aalto}
\affil{Onsala Rymdobservatorium, Chalmers Tekniska H\"ogskola,
S-439 92 Onsala, Sweden}
\author{S. H\"uttemeister}
\affil{Astronomisches Institut der Ruhr-Universit\"at Bochum, Universit\"atsstr. 150, 447 80 Bochum,
Germany}
\author{F. Walter}
\affil{Division of Physics, Mathematics and Astronomy, Caltech 105-24, Pasadena CA 91125, USA}

\begin{abstract}

High resolution observations of \twco\ and \thco\ 1--0 in the Medusa (NGC~4194) minor
merger show the 
${{^{12}{\rm CO}} \over {^{13}{\rm CO}}}$ 1--0 intensity ratio (\r) increasing
from normal values (5-10) in the outer parts of the galaxy to high ($>$ 20) values in the
central, extended starburst region. Ratios $>20$ are otherwise typical of more luminous 
mergers.
The Medusa ${L_{\rm FIR}\over L_{\rm CO}}$ ratio rivals that of ultraluminous galaxies (ULIRGs),
despite the comparatively modest luminosity, indicating an exceptionally high star formation 
efficiency (SFE). 
We present models of the high pressure ISM in a ULIRG and the relatively
low pressure ISM of the Medusa. We discuss how these models may explain large
\r\ in both types of distributions. Since the HCN emission is faint towards the Medusa, we suggest
that the SFE is not primarily controlled by the mass fraction of dense ($n \gapprox 10^4$$\cmmd$) gas,
but is probably strongly dependent on dynamics.  The bright HCN emission towards
ULIRGs is not necessarily evidence that the IR emission there is always powered by starbursts.
\end{abstract}

\section{Introduction}

Infrared luminous ($L_{\rm IR} > 10^{11}$ L$_{\odot}$) and ultraluminous galaxies
(ULIRGs) ($L_{\rm IR} > 10^{12}$ L$_{\odot}$) are believed to be major mergers of massive,
gas-rich disk galaxies. They are spectacular systems, related to super-starbursts
and massive gas flows to their central regions. Their molecular gas distribution
is often found in rotating nuclear disks, toruses or bars only a few
hundred pc in extent. 
These ULIRGs have been suggested as precursors of QSOs (e.g.
Norman and Scoville 1988). Still, galaxies collide that do not have the necessary properties
to become ultraluminous objects. The importance of these intermediate, often minor i.e. unequal-mass mergers,
to the general evolution of galaxies is not well understood.
Both major and minor mergers undergo bursts of star formation, but the activity seems
to take place on a larger linear scale in the minor mergers. Whether the starburst processes and
triggering mechanisms remain the same for the compact and large scale bursts is unclear.
It appears however, that some of the extended bursts can reach the same high SFEs as the ULIRGs.

Probing the properties and distribution of the molecular ISM enables modelling of
the triggering and evolution of the starburst.
For this purpose, it is necessary
to go beyond \twco\ and use other (fainter) molecular tracers such as the isotopomer
\thco\ and high dipole moment molecules such as HCN, HNC and CN. These lines are powerful
diagnostic tools, in particular in combination with other extinction-free tracers like IR and
radio.

\begin{figure}
\plottwo{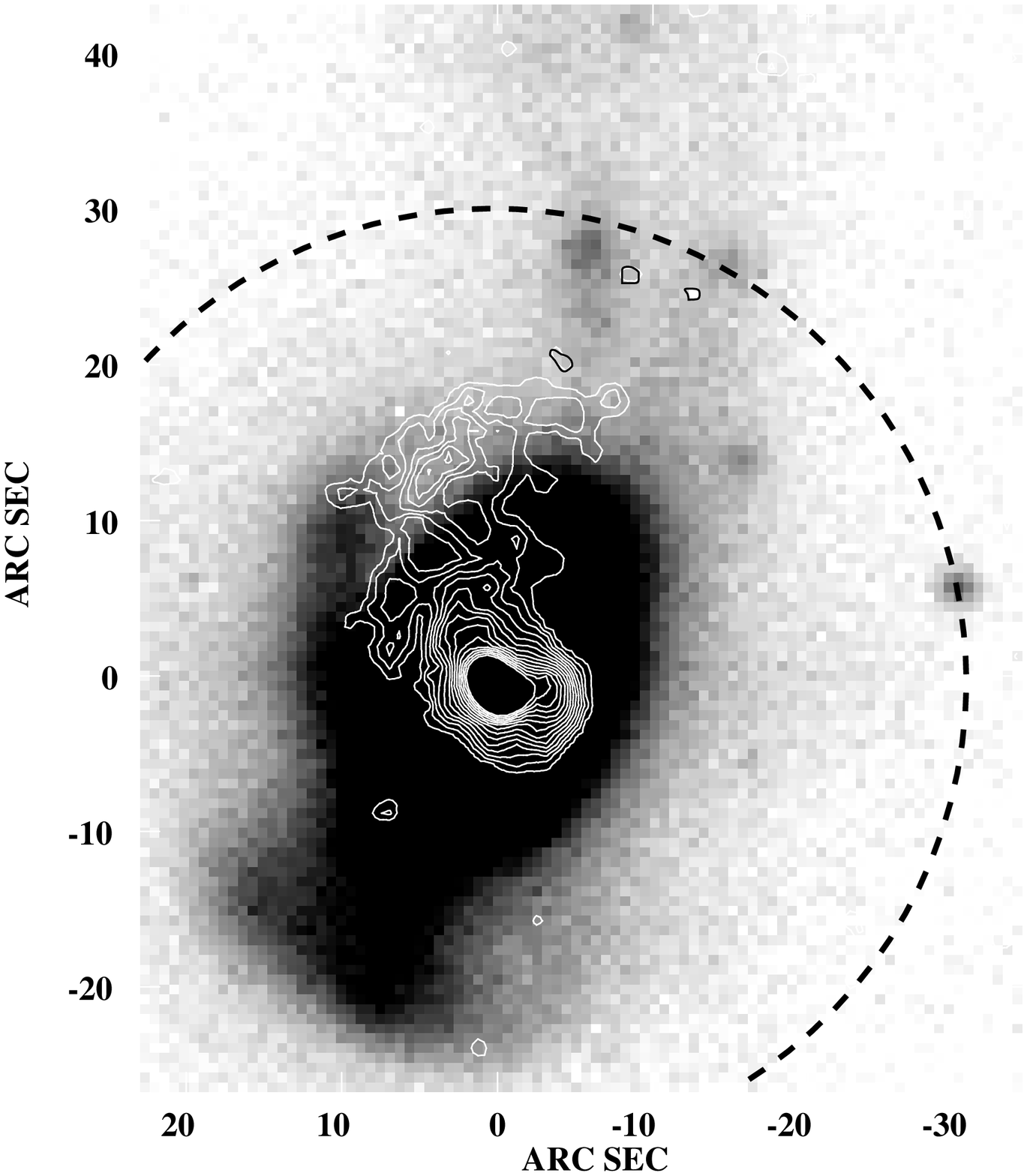}{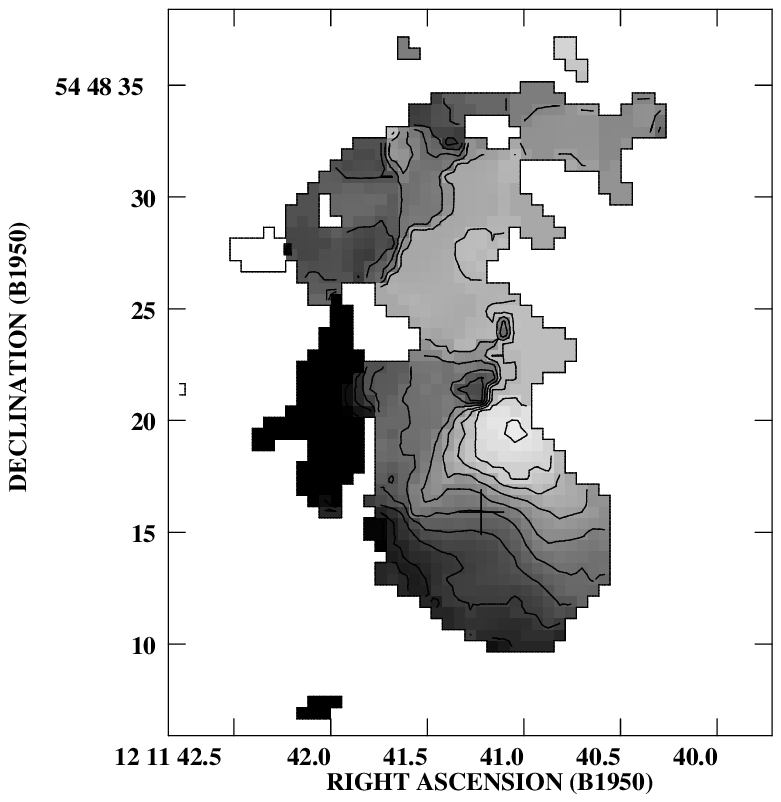}
\caption{The left panel shows an overlay of the \twco\ (not primary beam corrected) contours (white, apart from clouds in the
tail marked in black) on a greyscale, overexposed
optical R-band image (Mazzarella \& Boroson 1993). The dashed curve marks the edge of the
OVRO primary beam. The right panel shows the \twco\ velocity field. Contours range from 2400
to 2640 \kms, with spacing 24 \kms. The grayscale ranges from 2400 (light) to 2650 (dark) \kms.
The radio continuum peak is marked with a cross. Note that the position angle of the \twco\
distribution is perpendicular to the major axis of the velocity field because of the crossing
of the central dust lane.}
\end{figure}

The minor merger, and young shell galaxy, NGC~4194 (``the Medusa'') has an extended
region (2 kpc) of intense star formation (e.g. Prestwich \etal\ 1994)
which is responsible for most of the FIR luminosity ($L_{\rm IR} = 8.5 \times 10^{10}$
L$_{\odot}$ at $D$=39 Mpc). The SFE, ${L_{\rm FIR} \over \Mgs}$,
is high, 40 $\Lsun\,\Msun^{-1}$, {\it similar to the SFE of the ULIRG Arp~220}. The global
\r\ is large, $\approx 20$, indicating a highly excited or
disrupted ISM (Aalto \& H\"uttemeister 2000, AH). Despite the high SFE, emission from the high density tracer molecule HCN
remains undetected in NGC~4194. 
We present a preliminary analysis of the \twco\ and \thco\ distribution obtained at high
resolution with OVRO. We offer an explanation for the apparent lack of dense gas in the Medusa and
suggest ``cartoon models'' of a high pressure (ULIRG) and a comparatively low pressure (Medusa)
molecular ISM.

\section{The ${{^{12}{\rm CO}} \over {^{13}{\rm CO}}}$ 1--0 intensity ratios}

Since both high kinetic temperatures and large turbulent
line widths will decrease the \twco\ 1--0 optical depth ($\tau_{\rm CO}$), a map of the 
${{^{12}{\rm CO}} \over {^{13}{\rm CO}}}$ 1--0
intensity ratio (\r) can be used to identify regions
of extreme or unusual physical conditions in the molecular gas. Large ratios indicate low to moderate $\tau_{\rm CO}$
and Aalto \etal\ (1995) established
some general diagnostics of the cloud conditions and environment based on global values of \r: 
Small ratios, \r$\approx 6$ are an indication
of a normal Galactic disk population of clouds dominated by cool giant molecular clouds (GMCs); intermediate ratios
$10 \lapprox {\cal R}_{\rm 1-0} \lapprox 15$ are associated with the inner regions of normal starburst
galaxies; the extreme values ${\cal R}_{\rm 1-0} > 20$ originate in turbulent, {\it high pressure} gas in 
the centers of luminous mergers. In the most luminous mergers, the gas surface density implied by
the Galactic conversion factor from \twco\ luminosity to M(H$_2$) is well over
10$^{4}$~\Msun\ pc$^{-2}$---two orders of magnitude higher than in typical Milky Way 
GMCs. This led Aalto \etal\ (1995) to suggest that large values of \r\ are related
to the large gas surface densities in compact nuclear starbursts. Large surface densities require high
pressures in hydrostatic equilibrium; as the \twco\ ${2-1 \over 1-0}$ line ratios 
in these objects indicate {\it low} ($n < 10^3$ $\cmmd$) densities of the \twco\ emitting
gas-component it must be supported by large
turbulent line widths ($P \propto n (\delta V)^2$).  Thus, $\tau_{\rm CO}$ can be reduced to moderate
($\approx 1$) values, resulting in large \r. 
However, the global \r\ of the Medusa is high,
but the gas surface density is more than an order of magnitude 
lower than that of the ULIRGs. Therefore, the pressure will be considerably lower and the ISM is
likely much less turbulent (even though it is still significantly higher than the average
pressure for the Galactic disk). Thus, for the Medusa, the high \r\ may well be caused by elevated kinetic
temperatures rather than large line widths. There is indeed a strong
correlation between FIR colour temperature and 
large values of \r: Galaxies with ${\cal R}_{\rm 1-0} > 20$ all have
$f(60 \mu{\rm m})/f(100 \mu{\rm m})$ flux ratios $\gapprox 0.8$ (e.g. Aalto \etal\ 1991, 1995)
indicating high average dust temperatures.
Thus, large values of \r\ are also related to the effect of the starburst itself, heating dust and gas to high
temperatures.

\subsection{\twco\ and \thco\ morphology of NGC~4194}

Figure 1 shows the \twco\ emission overlayed on an R-band image of NGC~4194. The
\twco\ distribution is surprisingly extended, $\approx 5$ kpc, for an advanced merger 
even if the bulk (60 - 70 \%) of the emission emerges from the 2 kpc starburst region. The morphology
is complex, tracing large-scale dust lanes, one which curves along the north-eastern edge of
the main body and continues into the tidal tail, and one which is
crossing the central region along the minor axis. The gas mass (for a standard CO to H$_2$
conversion factor) is $2 \times 10^9$ M$_{\odot}$ (AH).
There is substantial variation in \r\ across NGC~4194:
From quiescent values of 7 in the the eastern dust lane 5 kpc from the center, to high values, $>$20,
in the starburst and central dust lane. This clearly demonstrates that there is a strong connection between
\r\ and gas environment.
In Figure 2 we show the \twco\ and \thco\ contours overlayed on an archival HST WFPC2 image of the inner 2 kpc
of NGC~4194. The \twco\ and \thco\ morphologies differ substantially showing significant variation in \r\
also within the central region. \thco\ peaks west of \twco, associated with a peculiar dust feature.
Elevated values of \r\ are found in two regions: a) througout the extended starburst region and b) in the 
curved part of the central dust lane. The global \r\ is dominated by the starburst region where \r\ is likely elevated
because of large kinetic temperatures (see next section). In region b) gas streaming may cause local effects of
large line widths (see also H\"uttemeister and Aalto, this volume).

\begin{figure}
\plotone{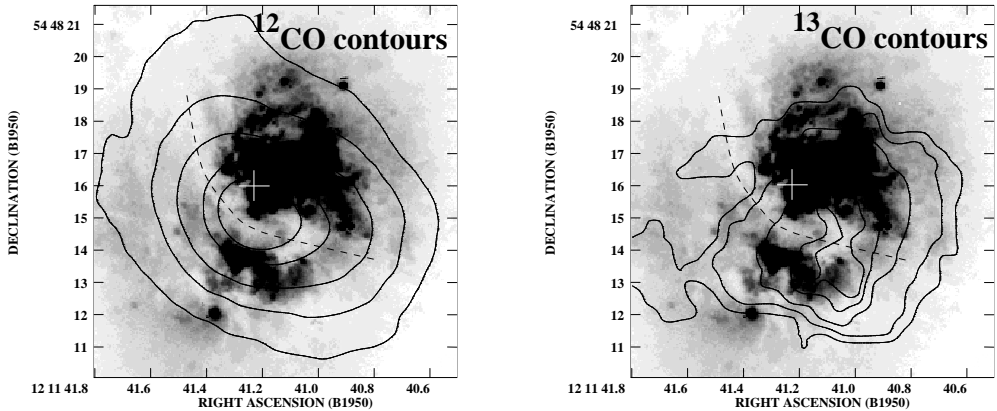}
\caption{Left image shows CO contours overlayed on HST WFPC2 image, and the right
shows the \thco\ 1--0 contours. The contours for both images have levels 
(20,40,60,80,90)\% of the peak flux. The resolution is 4$''$  for both CO and \thco.
The dashed line outlines the dust lane and the cross the radio continuum peak.}
\end{figure}

\begin{table}
\caption{Line Ratios in high \r\ starbursts}
\begin{tabular}{lcc}
Line & High pressure (compact) & Low pressure (extended) \\
\hline
CO 2-1/1-0 & subthermal ($\lapprox 0.6$) & thermalized ($\approx 1$) \\
${\cal R}_{\rm 1-0}$ & $\gapprox 20$ & $\gapprox 20$ \\
CO/HCN & 8 & $\gapprox 25$ \\

\end{tabular}
\end{table}

\subsection{The molecular ISM of an extended and compact starburst}

In the table we list typical line ratios for compact high pressure and extended low pressure
starbursts. Both types may have large values of ${\cal R}_{\rm 1-0}$,
but their average ISM properties differ on several accounts. In Figure 3 we show cartoons
of the two ISM types, ``raisin roll'' and ``fried eggs''.
For the former (ULIRG) scenario, the \twco\ emission is emerging from low density $n \lapprox 10^3$ $\cmmd$ 
diffuse gas of large linewidths and filling factor, while the HCN 1-0 emission is coming from
dense ($n \gapprox 10^4$ $\cmmd$), embedded clouds (but see e.g. Aalto \etal\ 1995 for a
discussion on mid-IR pumping of HCN). The bulk of the molecular mass is in the dense gas, but
the high value of \r\ can primarily be attributed to the large linewidths of the diffuse molecular gas (which can remain molecular
because of the extreme pressure). Galaxies which may be characterised by this ISM scenario include
Arp~220, NGC~6240 and Mrk~231.

It may seem surprising that the \twco\ emitting gas is somewhat denser,
$10^3 \lapprox n \lapprox 5 \times 10^3$ $\cmmd$, in the lower pressure scenario
(and with a smaller volume filling factor). This may be because 
(in contrast to the ``raisin roll'' ISM) the intercloud medium is likely atomic (or ionized)
since the destructive forces of the newborn stars are not balanced by a high ambient pressure. 
Thus, the \twco\ emitting gas is more ``cloudy'' in the lower pressure model, but the relatively faint HCN
emission indicates that the mass fraction of {\it high} density gas, $n > 10^4$ $\cmmd$, is considerably
lower.
The \twco\ emission may, however,
be filamentary and not always in bound clouds. The large value of \r\ is
likely caused by elevated gas temperatures ($T_{\rm k} \gapprox 50 $ K) since the 
gas is dense enough for \twco\ to be thermalized ($T_{\rm ex} \approx T_{\rm k}$). The clouds may be
PDRs (photon dominated regions) if the impact of the starburst is strong enough. 
Galaxies which may be characterised by this ISM scenario include
the Medusa (NGC~4194), UGC~2866 and NGC~1614.

\begin{figure}
\plottwo{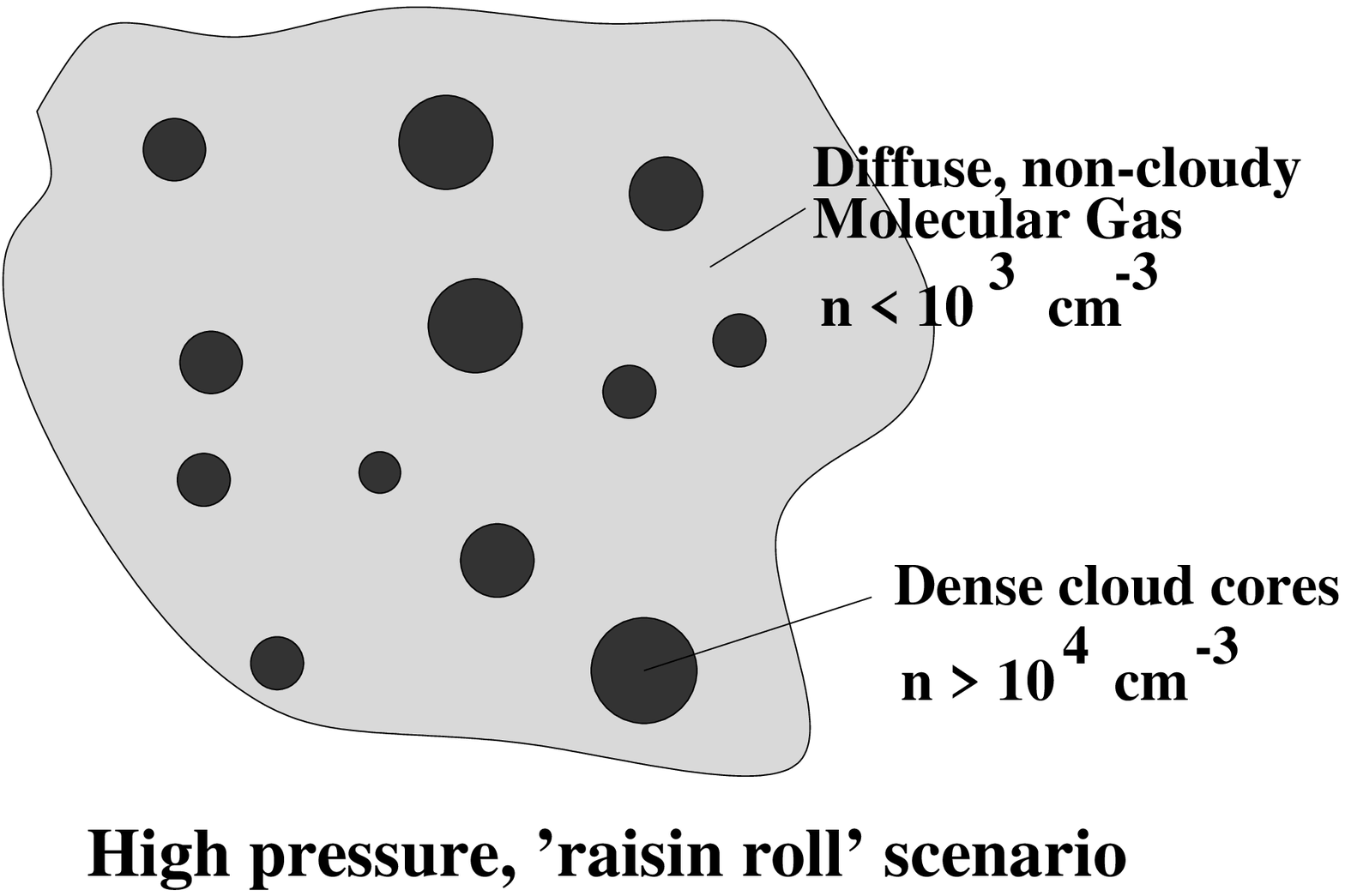}{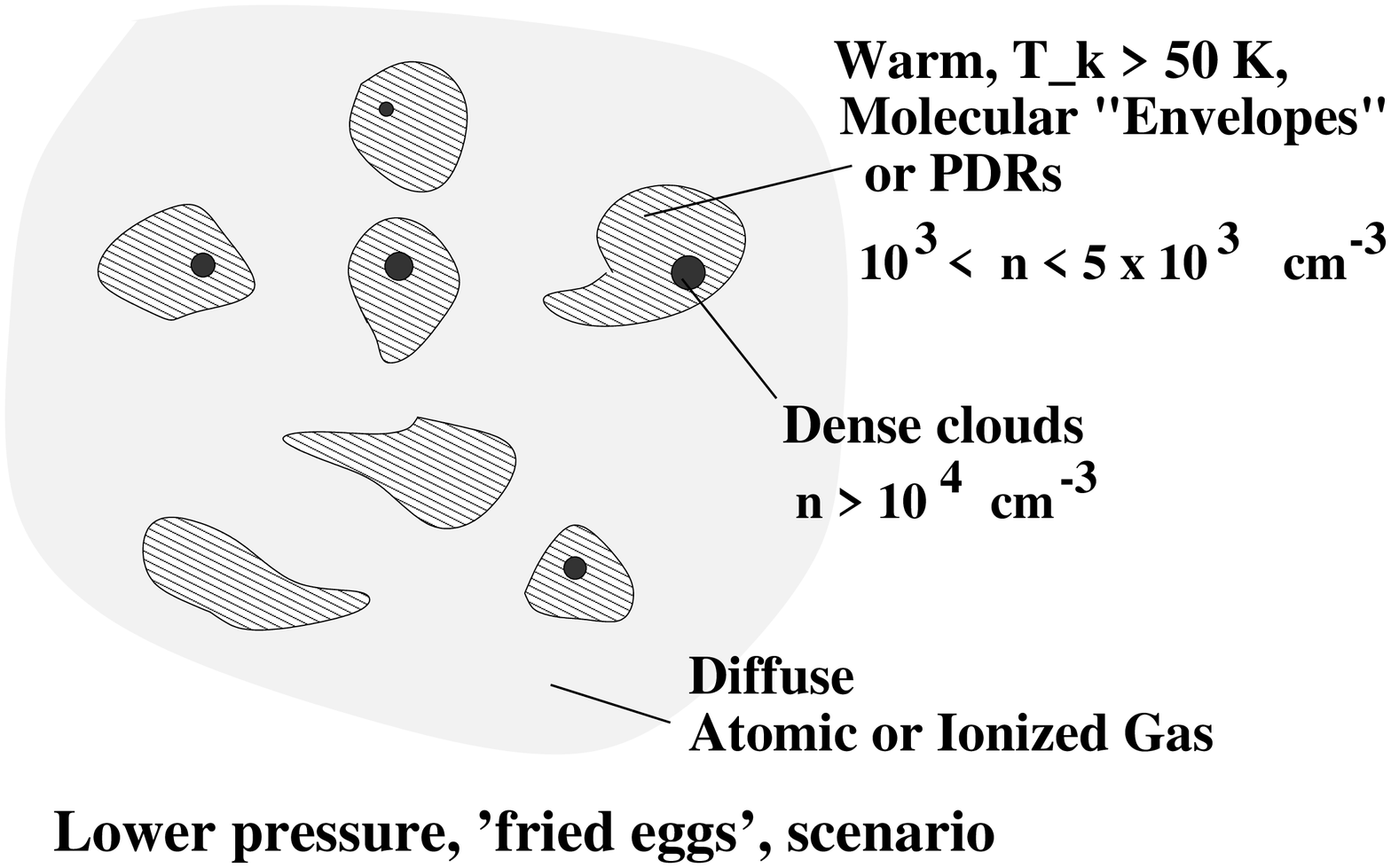}
\caption{Cartoon models of a high pressure and low pressure ISM}
\end{figure}

\section{HCN --- tracing star formation or pressure?}

Because of its high dipole moment, HCN 1-0 emission is often used as a tracer of high
density ($n > 10^4$ $\cmmd$) gas and the ${{^{12}{\rm CO}} \over {\rm HCN}}$ line intensity ratio as a
measure of the mass fraction of
dense gas. Bright HCN (low ${{^{12}{\rm CO}} \over {\rm HCN}}$ ratios) in ULIRGs has been used to argue for
a starburst origin of their luminosity (e.g. Solomon, Downes and Radford 1992)
because of a global correlation between $L_{\rm HCN}$ and $L_{\rm FIR}$.
It is therefore interesting that the SFE of the Medusa rivals that of the ULIRG 
Arp~220 --- even though the HCN emission is faint towards NGC~4194.  How
can there be an efficient transformation of gas into stars when there is only little dense gas present?
The answer may lie in the time the molecular gas spends in a dense
phase which is related to the average gas pressure and dynamics.
The Medusa starburst takes place in an environment of reduced shear because it occurs inside the 
region of solid body rotation. Thus, gravitational instability may dominate over tidal shear
with a resulting increase in the SFE. The rate of star formation (SFR) is indeed very high,
40 M$_{\odot}$ yr$^{-1}$, which is close to the maximum SFR per kpc$^2$ found by Lehnert \& Heckman (1996).
Since the hydrostatic pressure is much lower than in a ULIRG, gas is not
maintained at high densities, but passes through a dense phase quickly on its way to becoming stars.
The time spent in a dense phase is short and will not be an observed signature of the ISM.
The rate of star formation has been found to be dependent on large scale dynamics (e.g. Kenney, Carlstrom and Young 1993).
Perhaps, for a threshold level of the gas surface density, the dynamical
environment of the clouds is also important in regulating the {\it efficiency} of star formation, even
more so than the relative amount of high density gas.
 
For the compact burst in the deep potential well of a massive major merger,  the
gas surface density (and matter density) is high and thus also the ambient pressure
leading to large {\it average} gas densities. Still, the dynamical and radiative environment
may be unfavourable to star formation.
Within 200 pc of the deep potential well of a differentially rotating galaxy with rotational
speed 250 \kms, the clouds must have average densities
of $n > 10^4$ $\cmmd$ just to be stable against tidal shearing. Lower density gas will only
exist as diffuse unbound clouds. 
In addition, the feedback mechanisms of the starburst itself may help regulate the SFE, and these mechanisms may
be effective in the densely packed, high gas surface density central regions of ULIRGs. 
Thus, in some circumstances, the gas will remain at substantial
average density without forming stars. In this context, the $L_{\rm HCN}$ - $L_{\rm FIR}$ correlation is,
at least partially, caused by high pressure gas
being more centrally concentrated than low pressure gas, and that FIR emission in galaxies
tend to emerge from the inner regions --- at least the 60 $\mu$m emission. The FIR emission
may indeed originate in starburst activity, but also from an AGN or from an evolved starburst
where densely packed stars heat a fragmented ISM.

A possible caveat in the notion that the SFE is similar in the Medusa and Arp~220 
is the possibility that the \twco\ luminosity is tracing molecular mass differently in a compact
and extended starburst. Furthermore, the ULIRGs deviate from the limiting SFR per kpc$^2$ found
by Lehnert and Heckman. Their maximum star formation rates
are in the range 100-300 M$_{\odot}$ yr$^{-1}$ instead of 20-40 --- and from a smaller area than less powerful
galaxies. Dust opacity and hidden AGNs may explain some of the discrepancy in the limiting SFRs, but we cannot
exclude that the underlying star formation mechanisms are different.

%\begin{quote}

%\end{quote}


\begin{references}

%\reference Aalto, S., H\"uttemeister, S., \& Polatidis, A.G. 2001, A\&A Letters, in press

\reference Aalto, S., \& H\"uttemeister, S. 2000,  A\&A, 362, 42 (AH)

\reference Aalto, S., Booth, R.S., Black, J.H., Johansson, L.E.B. 1995, A\&A, 300, 369

\reference Aalto, S., Black, J.H., Johansson, L.E.B., Booth, R.S. 1991, A\&A, 249, 323

%\reference Casoli, F., Dupraz, C., Combes, F. 1992, A\&A 264, 55

\reference Kenney, J.D.P., Carlstrom, J.E., Young, J.S. 1993, ApJ, 418, 687

\reference Lehnert, M.D., \& Heckman, T.M. 1996, ApJ, 472, 546

\reference Mazzarella J.M., Boroson T.A. 1993, ApJS 85, 27

\reference Norman, C., and Scoville, N., 1988, ApJ, 332, 124

\reference Prestwich A.H., Joseph R.D., Wright G.S. 1994, ApJ 422, 73

\reference Solomon P.M., Downes D., \& Radford S.J.E. 1992, ApJ 387, L55 

\end{references}
\end{document}